\documentclass[aps,prl,twocolumn,superscriptaddress,showpacs]{revtex4}

\usepackage{amsmath,amssymb,amsfonts,amsthm,url,xspace,graphicx,psfrag}
\usepackage[pdftitle={Driving sandpiles to criticality and beyond},
            pdfauthor={Anne Fey, Lionel Levine, David B. Wilson},
            bookmarks=true,bookmarksopen=true,bookmarksopenlevel=2,
            hyperfootnotes=false]{hyperref}
\usepackage[all]{hypcap}
\usepackage{multirow}

\newcommand{\arXiv}[1]{\href{http://arxiv.org/abs/#1}{arXiv:#1}}
\newcommand{\arxiv}[1]{\href{http://arxiv.org/abs/#1}{#1}}
\newcommand{\sref}[1]{\hyperref[#1]{\S~\ref*{#1}}}
\newcommand{\aref}[1]{\hyperref[#1]{Appendix~\ref*{#1}}}
\newcommand{\lref}[1]{\hyperref[#1]{Lemma~\ref*{#1}}}
\newcommand{\tref}[1]{\hyperref[#1]{Theorem~\ref*{#1}}}
\newcommand{\cref}[1]{\hyperref[#1]{Corollary~\ref*{#1}}}
\newcommand{\fref}[1]{\hyperref[#1]{Figure~\ref*{#1}}}
\newcommand{\pref}[1]{\hyperref[#1]{Proposition~\ref*{#1}}}
\renewcommand{\sref}[1]{\hyperref[#1]{section~\ref*{#1}}}

\newcommand{\old}[1]{}

\DeclareMathSymbol{\E}{\mathbin}{AMSb}{"45}
\DeclareMathSymbol{\EE}{\mathbin}{AMSb}{"45}
\DeclareMathSymbol{\Z}{\mathbin}{AMSb}{"5A}

\newtheorem{theorem}{Theorem}

\begin{document}

\title{Driving Sandpiles to Criticality and Beyond}

\author{\href{http://dutiosc.twi.tudelft.nl/~anne/}{Anne Fey}}
\affiliation{Delft Institute of Applied Mathematics, Delft University of Technology, The Netherlands}
\author{\href{http://math.mit.edu/~levine}{Lionel Levine}}
\affiliation{Department of Mathematics, Massachusetts Institute of Technology, Cambridge, MA 02139, USA}
\author{\href{http://dbwilson.com}{David B. Wilson}}
\affiliation{Microsoft Research, Redmond, WA 98052, USA}

\date{December 16, 2009; revised March 4, 2010}

\begin{abstract}
A popular theory of self-organized criticality relates driven dissipative systems to systems with conservation.  This theory predicts that the stationary density of the abelian sandpile model equals the threshold density of the fixed-energy sandpile.  We refute this prediction for a wide variety of underlying graphs, including the square grid.  Driven dissipative sandpiles continue to evolve even after reaching criticality.
This result casts doubt on the validity of using fixed-energy sandpiles to explore the critical behavior of the abelian sandpile model at stationarity.
\end{abstract}

\maketitle


In a widely cited series of papers
\cite{absorbing1,absorbing2,absorbing3,absorbing4,absorbing5},
Dickman, Mu{\~n}oz, Vespignani, and Zapperi (DMVZ) developed a theory of
self-organized criticality as a relationship between driven
dissipative systems and systems with conservation.  This theory
predicts a specific relationship between the abelian
sandpile model of Bak, Tang, and Wiesenfeld \cite{BTW}, a driven system
in which particles added at random dissipate across the boundary, and
the corresponding ``fixed-energy sandpile,'' a closed system in which
the total number of particles is conserved.

After defining these two models and explaining the conjectured
relationship between them in the DMVZ paradigm of self-organized
criticality, we present data from large-scale simulations which
strongly indicate that this conjecture is false on the two-dimensional
square lattice.  We then examine the conjecture on some simpler
families of graphs in which we can provably refute it.

Early experiments \cite{GM} already identified a discrepancy, at
least in dimensions 4 and higher, but later work focused on dimension
2 and missed this discrepancy (it is very small).  Some recent papers
(e.g., \cite{BM}) restrict their study to stochastic sandpiles because
deterministic sandpiles belong to a different universality class, but
there remains a widespread belief in the DMVZ paradigm for both
deterministic and stochastic sandpiles \cite{VD,CVSD}.

Despite our contrary findings,
we believe that the central idea of the DMVZ paradigm is
a good one: the dynamics of a driven dissipative system should in some
way reflect the dynamics of the corresponding conservative system.
Our results point to a somewhat different relationship than that
posited in the DMVZ series of papers: the driven dissipative model
exhibits a second-order phase transition at the threshold density of
the conservative model.  

Bak, Tang, and Wiesenfeld~\cite{BTW} introduced the abelian sandpile as
a model of self-organized criticality; for mathematical background,
see~\cite{frank}.  The model begins with a collection of particles on
the vertices of a finite graph.  A vertex having at least as many
particles as its degree \emph{topples\/} by sending one particle along
each incident edge.  A subset of the vertices are distinguished as
sinks: they absorb particles but never topple.  A single time step
consists of adding one particle at a random site, and then performing
topplings until each non-sink vertex has fewer particles than its degree.
The order of topplings does not affect the outcome~\cite{dhar}.
The set of topplings caused by addition of a particle is called an
avalanche.

Avalanches can be decomposed into a sequence of ``waves''
so that each site topples at most once during each wave.
Over time, sandpiles
evolve toward a stationary state in which the waves exhibit power-law
statistics \cite{KLGP} (though the full avalanches seem to exhibit
multifractal behavior \cite{MST,KMS}).  Power-law behavior is a
hallmark of criticality, and since the stationary state is reached
apparently without tuning of a parameter, the model is said to be
\emph{self-organized critical}.

To explain how the sandpile model self-organizes to reach the critical
state, Dickman \textit{et al.}~\cite{absorbing1, absorbing3}
introduced an argument which soon became widely accepted: see, for
example, \cite[Ch.~15.4.5]{sornette} and \cite{quant,feyredig,RS}.
Despite the apparent lack of a free parameter, they argued, the
dynamics implicitly involve the tuning of a parameter to a value where
a phase transition takes place.  The phase transition is between an
active state, where topplings take place, and a quiescent
``absorbing'' state.  The parameter is the \emph{density}, the average
number of particles per site.  When the system is quiescent, addition
of new particles increases the density.  When the system is active,
particles are lost to the sinks via toppling, decreasing the
density. The dynamical rule ``add a particle when all activity has
died out'' ensures that these two density changing mechanisms balance
one another out, driving the system to the threshold of instability.

To explore this idea, DMVZ introduced the
\emph{fixed-energy sandpile\/} model (FES), which involves an explicit
free parameter $\zeta$, the density of particles.  On a graph with $N$
vertices, the system starts with $\zeta N$ particles at vertices
chosen independently and uniformly at random.  Unlike the driven
dissipative sandpile described above, there are no sinks and no
addition of particles.  Subsequently the system evolves through
toppling of unstable sites. Usually the parallel toppling order is
chosen: at each time step, all unstable sites topple simultaneously.
Toppling may
persist forever, or it may stop after some finite time.  In the latter
case, we say that the system \emph{stabilizes}; in the terminology of
DMVZ, it reaches an ``absorbing state.''

A common choice of underlying graph for FES is the $n\times n$ square grid
with periodic boundary conditions.  It is believed, and supported by
simulations \cite{stairs}, that there is a \emph{threshold density\/}
$\zeta_c$, such that for $\zeta<\zeta_c$, the system stabilizes with
probability tending to~$1$ as $n \to \infty$; and for $\zeta>\zeta_c$,
with probability tending to~$1$ the system does not stabilize.

\section{The Density Conjecture}

For the driven dissipative sandpile on the $n\times n$ square grid
with sinks at the boundary, as $n \to \infty$ the stationary measure
has an infinite-volume limit \cite{AJ}, which is a measure on
sandpiles on the infinite grid $\Z^2$.  One gets the same limiting
measure whether the grid has periodic or open boundary conditions, and
whether there is one sink vertex or the whole boundary serves as a
sink \cite{AJ} (see also \cite{pemantle} for the corresponding result
on random spanning trees).  The statistical properties of this
limiting measure have been much studied
\cite{MD,priezzhevheights,JPR}.  Grassberger conjectured that the expected
number of particles at a fixed site is $17/8$,
and it is now known to be $17/8\pm 10^{-12}$ \cite{JPR}.
We call this value the \emph{stationary density\/} $\zeta_s$ of $\Z^2$.

DMVZ believed that the combination of driving and dissipation in the
classical abelian sandpile model should push it toward the threshold
density $\zeta_c$ of the fixed-energy sandpile.  This leads to a
specific testable prediction, which we call the Density Conjecture.

\vspace{2pt}
\noindent
\textbf{Density Conjecture \cite{absorbing4}.}
On the square grid, $\zeta_c = 17/8$.  More generally, $\zeta_c=\zeta_s$.
\vspace{2pt}

Vespignani \textit{et al.}~\cite{absorbing4} write of FES on the square grid,
``the system turns out to be critical
only for a particular value of the energy density equal to that of the
stationary, slowly driven sandpile.''  They add that the threshold
density $\zeta_c$ of the fixed energy sandpile is ``the only possible
stationary value for the energy density'' of the driven dissipative
model.  In simulations they find $\zeta_c = 2.1250(5)$, adding in a
footnote ``It is likely that, in fact, 17/8 is the exact result.''
Other simulations to estimate $\zeta_c$
also found the value very close to $17/8$ \cite{absorbing1,absorbing2}.


Our goal in the present paper is to demonstrate that the density
conjecture is more problematic than it first appears.
Table~\ref{table:Z2} presents data from large-scale simulations
indicating that $\zeta_c(\Z^2)$ is $2.125288$ to six decimals; close
to but not exactly equal to $17/8$.

In each trial, we added particles one at a time at uniformly random
sites of the $n\times n$ torus.  After each addition, we performed
topplings until either all sites were stable, or every site toppled at
least once.  For deterministic sandpiles on a
connected graph, if every site topples at least once, the system will
never stabilize \cite{tardos,FMR,FLP}.  We recorded $m/n^2$ as an empirical
estimate of the threshold density $\zeta_c(\Z_n^2)$, where~$m$ was the
maximum number of particles for which the system stabilized.  We
averaged these empirical estimates over many independent trials.

We used a random number generator based on the Advanced Encryption
Standard (AES-256), which has been found to exhibit excellent
statistical properties \cite{HW:AES,TestU01}.
Our simulations
were conducted on a High Performance Computing (HPC) cluster of
computers.

\newcommand{\stddev}{0.0000004}

\begin{table}
\begin{center}
\parbox{\columnwidth}{
\scriptsize
\begin{minipage}[b]{1.75in}
\begin{ruledtabular}
\begin{tabular}{rcc}
 $n$    &   trials & estimate of $\zeta_c(\Z_n^2)$ \\
\hline
   $64$ &   $2^{28}$ & $2.1249561 \pm \stddev$\\
  $128$ &   $2^{26}$ & $2.1251851 \pm \stddev$\\
  $256$ &   $2^{24}$ & $2.1252572 \pm \stddev$\\
  $512$ &   $2^{22}$ & $2.1252786 \pm \stddev$\\
 $1024$ &   $2^{20}$ & $2.1252853 \pm \stddev$\\
 $2048$ &   $2^{18}$ & $2.1252876 \pm \stddev$\\
 $4096$ &   $2^{16}$ & $2.1252877 \pm \stddev$\\
 $8192$ &   $2^{14}$ & $2.1252880 \pm \stddev$\\
$16384$ &   $2^{12}$ & $2.1252877 \pm \stddev$\\
\end{tabular}
\end{ruledtabular}
\end{minipage}
\hfill
\psfrag{n}[cr][cr]{$n$}
\psfrag{zc(Zn)}[bl][cl]{$\zeta_c(\Z_n^2)$}
\psfrag{2.125288}[Bl][Bl]{$2.125288$}
\psfrag{2.125}[Bl][Bl]{$2.125000000000$}
\includegraphics[width=0.45\columnwidth]{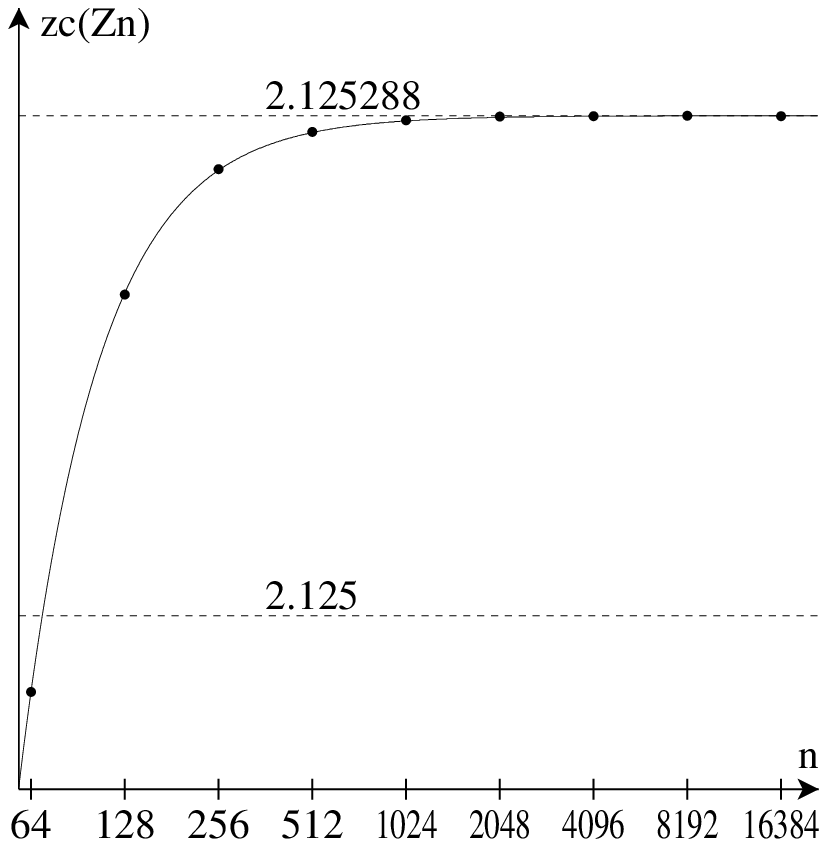}
}
\end{center}
\caption{
  Fixed-energy sandpile simulations on $n\times n$ tori~$\Z_n^2$.  The third column gives our empirical estimate of the threshold density~$\zeta_c(\Z_n^2)$ for $\Z_n^2$.  The standard deviation in each of our estimates of $\zeta_c(\Z_n^2)$ is $4\times 10^{-7}$.  To six decimals, the values of $\zeta_c(\Z_{2048}^2),\dots,\zeta_c(\Z_{16384}^2)$ are all the same.  The data from $n=64$ to $n=16384$ are well approximated by $\zeta_c(\Z_n^2) = 2.1252881\pm 3\times 10^{-7} - (0.390\pm 0.001) n^{-1.7}$, as shown in the graph.  (The error bars are too small to be visible, so the data are shown as points.)  The rapid convergence is due in part to periodic boundary conditions.  We conclude that the asymptotic threshold density $\zeta_c(\Z^2)$ is $2.125288$ to six decimals.  In contrast, the stationary density $\zeta_s(\Z^2)$ is $2.125000000000$ to twelve decimals.
}
\label{table:Z2}
\end{table}

\section{Phase transition at \texorpdfstring{$\zeta_c$}{threshold}}

We consider the density conjecture on several other families of
graphs, including some for which we can determine the exact
values~$\zeta_c$ and~$\zeta_s$ analytically.

Dhar \cite{dhar} defined recurrent sandpile configurations and showed
that they form an abelian group.  A consequence of his result is that
the stationary measure for the driven dissipative sandpile on a finite
graph $G$ with sinks is the uniform measure on recurrent
configurations.  The \emph{stationary density\/} $\zeta_s(G)$ is the
expected total number of particles in a uniform random recurrent
configuration, divided by the number of non-sink vertices in $G$.

The threshold density~$\zeta_c$ and
stationary density~$\zeta_s$ for different graphs is summarized in
Table~\ref{table:summary}.  The only graph on which the two densities
are known to be equal is $\Z$ \cite{quant,feyredig,FMR}.  On all other
graphs we examined, with the possible exception of the $3$-regular Cayley
tree, it appears that $\zeta_c \neq \zeta_s$.

Each row of Table~\ref{table:summary} represents an infinite family of
graphs~$G_n$ indexed by an integer~$n \geq 1$.  For example,
for~$\Z^2$ we take~$G_n$ to be the~$n \times n$ square grid, and for
the regular trees we take~$G_n$ to be a finite tree of depth~$n$.  As
sinks in~$G_n$ we take the set of boundary sites~$G_n \setminus
G_{n-1}$ (note that on trees this corresponds to wired
boundary conditions).  The value of~$\zeta_s$ reported is $\lim_{n \to \infty} \zeta_s(G_n)$.

\begin{table}[t!]
\begin{center}
\begin{ruledtabular}
\begin{tabular}{ccc}
graph & $\zeta_s$ & $\zeta_c$ \\
\hline
$\Z$ & {\bf 1} & {\bf 1} \\
$\Z^2$ & $\mathbf{17/8}=2.125$  & $2.125288\ldots$ \\
bracelet & $\mathbf{5/2} = 2.5$ &  $\mathbf{2.496608\ldots}$  \\
flower graph & $\mathbf{5/3} = 1.666667\ldots$ & $\mathbf{1.668898\ldots}$ \\
ladder graph & $\mathbf{\frac74 - \frac{\sqrt{3}}{12}} = 1.605662\ldots$ & $1.6082\ldots$ \\
complete graph & $\mathbf{1/2}\times n + O(\sqrt{n})$ & $\mathbf{1}\times n-O(\sqrt{n \log n})$ \\
3-regular tree & $\mathbf{3/2}$  & 1.50000\dots \\
4-regular tree & $\mathbf{2}$    & 2.00041\dots \\
5-regular tree & $\mathbf{5/2}$  & $2.51167\dots$ \\
\end{tabular}
\end{ruledtabular}
\end{center}
\caption{Stationary and threshold densities for different graphs.  Exact values are in bold.
}
\label{table:summary}
\end{table}

The exact values of $\zeta_s$ for regular trees (Bethe lattices)
were calculated by Dhar and Majumdar \cite{DM}.
The corresponding values of $\zeta_c$ we report come from simulations
\cite{FLW:approach}.
We derive or simulate the values of $\zeta_s$ and $\zeta_c$ for
the bracelet, flower, ladder, and complete graphs in \cite{FLW:approach}.

\begin{figure}[t]
\begin{center}
 \includegraphics[width=0.3\columnwidth]{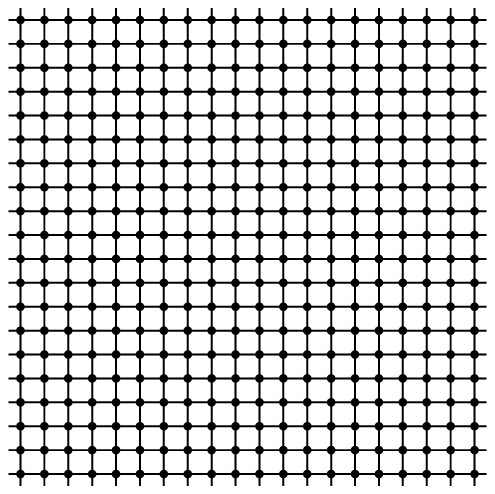}
 \hspace{0.1\columnwidth}
 \includegraphics[width=0.3\columnwidth]{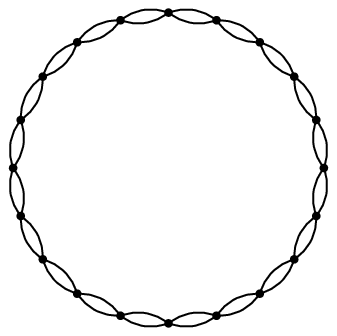}
\end{center}
\begin{center}
 \includegraphics[width=0.3\columnwidth]{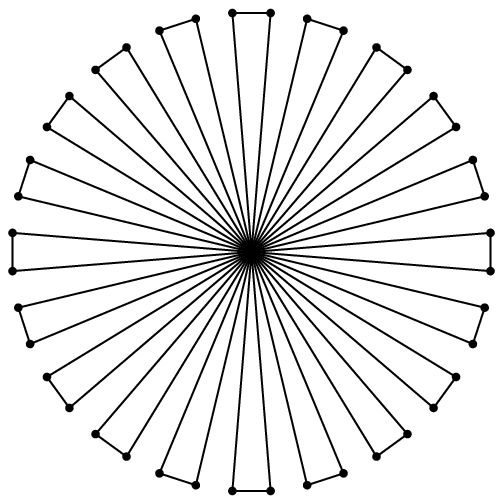}
 \hspace{0.1\columnwidth}
 \includegraphics[width=0.3\columnwidth]{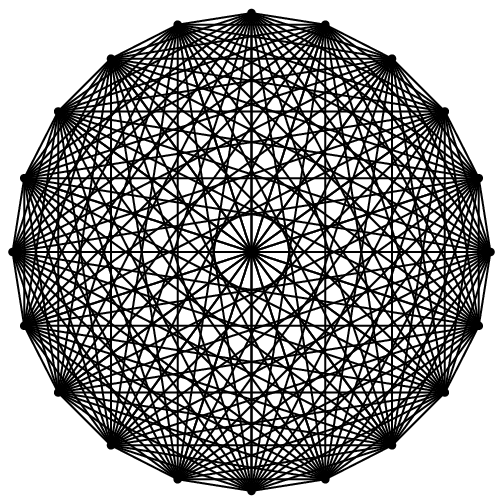}
\end{center}
\begin{center}
\includegraphics[width=\columnwidth]{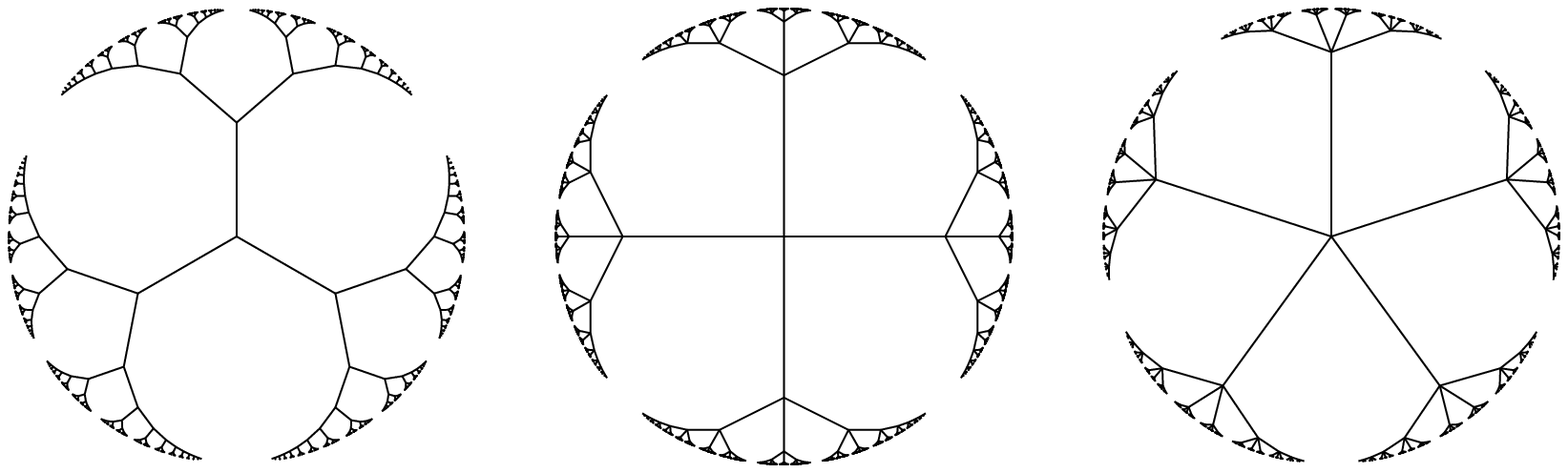}
\end{center}
\begin{center}
\includegraphics[width=0.8\columnwidth]{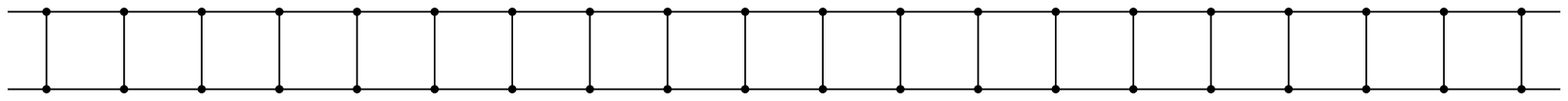}
\end{center}
\caption{The graphs on which we compare $\zeta_s$ and $\zeta_c$: the grid (upper left), bracelet graph (upper right), flower graph (2$^{\text{nd}}$ row left), complete graph (2$^{\text{nd}}$ row right), Cayley trees (Bethe lattices) of degree $d=3,4,5$ (3$^{\text{rd}}$ row), and ladder graph (bottom).}
\label{fig:graphs}
\end{figure}

As an example, consider the \emph{bracelet graph\/} $B_n$,
which is a cycle of $n$ vertices, with each edge doubled
(see Figure~\ref{fig:graphs}).  A site topples by
sending out $4$ particles: $2$ to each of its two neighbors.  One site
serves as the sink.  To compare the densities $\zeta_c$ and $\zeta_s$,
we consider the driven dissipative sandpile before it reaches
stationarity, by running it for time $\lambda$.  More precisely, we
place $\lambda n$ particles uniformly at random, stabilize the
resulting sandpile, and let $\rho_n(\lambda)$ denote the expected
density of 
the resulting stable configuration.
In the long version of this paper \cite{FLW:approach} we prove

\begin{theorem}[\cite{FLW:approach}]
\label{braceletmain}
For the bracelet graph
$B_n$, in the limit as $n\to \infty$,
\begin{enumerate}
\item The threshold density $\zeta_c$
is the unique positive root of $\zeta = \frac52 - \frac12 e^{-2\zeta}$
(numerically, $\zeta_c = 2.496608$).

\item The stationary density $\zeta_s$ is $5/2$.

\item
The final density $\rho_n(\lambda)$,
as a function of initial density $\lambda$,
converges pointwise
to a limit $\rho(\lambda)$, where
        \[ \rho(\lambda) = \min\left (\lambda, \frac{5-e^{-2\lambda}}{2}\right) =   \begin{cases} \lambda, & \lambda \leq \zeta_c \\ \frac{5-e^{-2\lambda}}{2}, & \lambda>\zeta_c. \end{cases} \]
\end{enumerate}
\end{theorem}

\begin{figure}
\psfrag{r}{$\rho$}
\psfrag{l}{$\lambda$}
\psfrag{rc}[Bc][Bc][0.8]{$\zeta_c$}
\psfrag{rs=5/2}[Bc][cc][0.8]{\hspace{30pt} $\zeta_s=5/2$}
\psfrag{curve}[Bc][Bc][0.8]{\hspace{12pt}$\frac{5-e^{-2\lambda}}{2}$}
\includegraphics[width=\columnwidth]{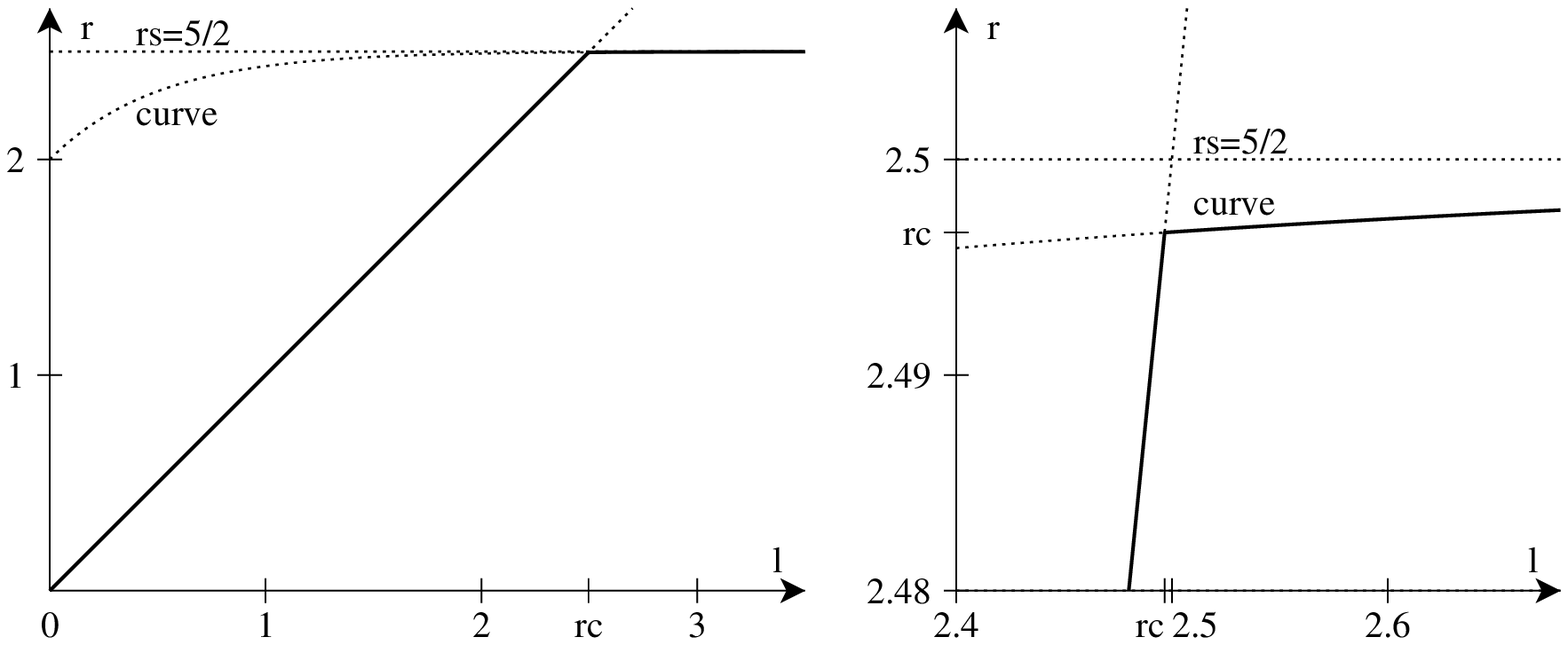}
\\[36pt]
\psfrag{rs=5/3}[cc][Bc][0.8]{\hspace{20pt}$\zeta_s=5/3$}
\psfrag{curve}[Bc][tc][0.8]{$\frac{5+e^{-3\lambda}}{3}$}
\includegraphics[width=\columnwidth]{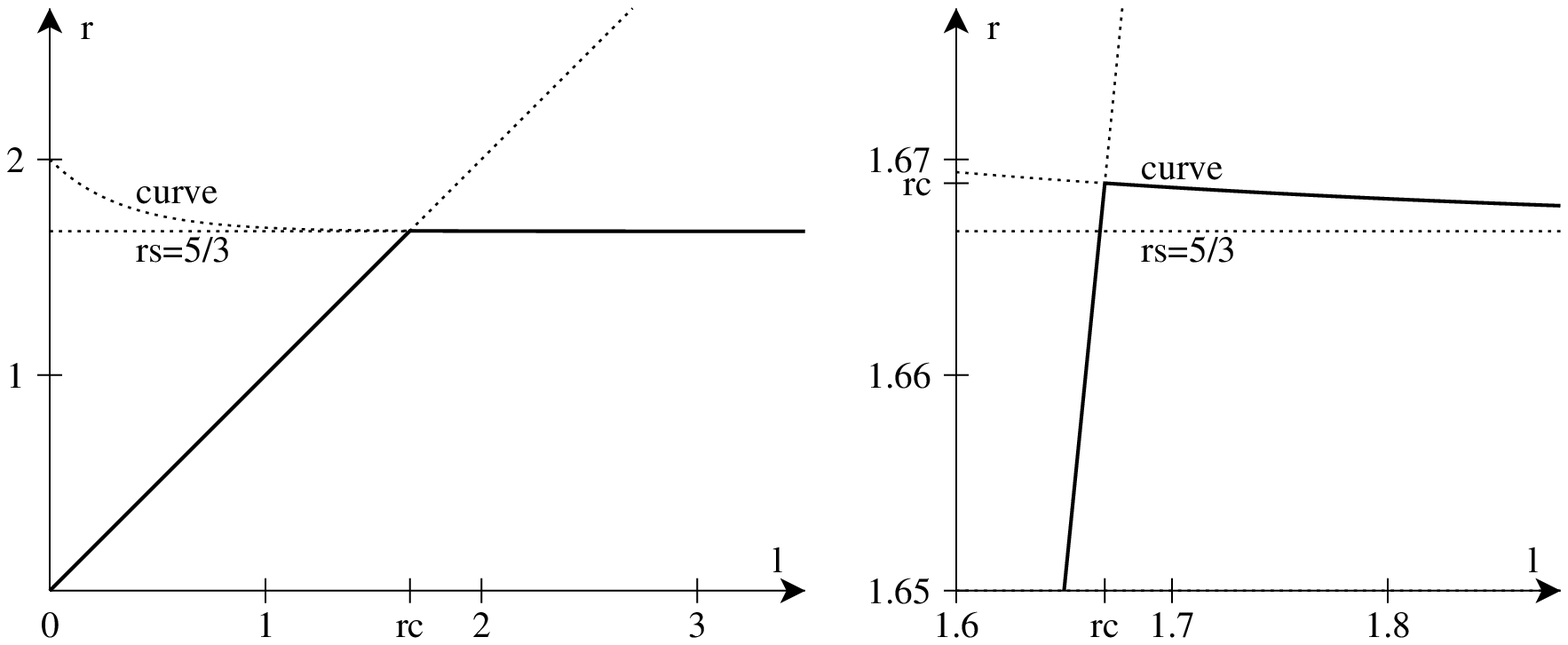}
\caption{Density $\rho(\lambda)$ of the final stable configuration as a function of initial density $\lambda$ on the bracelet graph (top row) and flower graph (bottom row) as the graph size tends to infinity.  A phase transition occurs at $\lambda=\zeta_c$.  At first glance (left panels) it appears that the driven sandpile reaches its stationary density $\zeta_s$ at this point, but closer inspection (right panels) reveals that the final density $\rho(\lambda)$ continues to change as $\lambda$ increases beyond~$\zeta_c$.  These graphs are exact.}
\label{fig:density}
\end{figure}

Part~3 of this theorem shows that despite the inequality $\zeta_s \neq
\zeta_c$, a connection remains between the driven dissipative dynamics
used to define $\zeta_s$ and the conservative dynamics used to define
$\zeta_c$: since the derivative $\rho'(\lambda)$ is discontinuous at
$\lambda=\zeta_c$, the driven sandpile undergoes a second-order phase
transition at density $\zeta_c$.
For $\lambda<\zeta_c$, the driven sandpile loses very few particles to
the sink, and the final density equals the initial density $\lambda$;
for $\lambda > \zeta_c$ it loses a macroscopic proportion of particles to
the sink, so the final density is strictly smaller than~$\lambda$.  As
Figure~\ref{fig:density} shows, the sandpile continues to evolve as
$\lambda$ increases beyond~$\zeta_c$; in particular, its density keeps
increasing.

We are also able to prove that a similar phase transition occurs on
the \emph{flower graph}, shown in Figure~\ref{fig:graphs}.
Interestingly, the final density $\rho(\lambda)$ for the flower graph
is a \emph{decreasing\/} function of $\lambda > \zeta_c$
(Figure~\ref{fig:density} bottom).

Our proofs make use of local toppling invariants on these graphs. On
the bracelet graph, since particles always
travel in pairs, the parity of the number of particles on a
single vertex is conserved. On the flower graph, the difference
modulo~$3$ of the number of particles on the two vertices in a single
``petal'' is conserved.

One might guess that the failure of the density conjecture is due only
to the existence of local toppling invariants, or else to boundary
effects from the sinks.  The ladder graph (Figure
\ref{fig:graphs}) has no local toppling invariants; moreover, it is
essentially one-dimensional, so the bulk of the graph is well
insulated from the sinks at the boundary.
Nevertheless, we find \cite{FLW:approach} a small but undeniable
difference between $\zeta_s$ and $\zeta_c$ on the ladder graph.
\old{
:  From the work of J\'{a}rai and Lyons~\cite{JL} and the
Parry formula~\cite{parry}, we compute $\zeta_s = \frac74 -
\frac{\sqrt{3}}{12}$ (numerically, $\zeta_s = 1.6057$), while our
large-scale simulations ($2^{12}$ trials with $n=2^{18}$) indicate
that $\zeta_c$ is about $1.6082$.
}

\old{
\section{Complete Graph}

The minimum density of a recurrent configuration on the complete graph $K_n$ is $\frac{n-2}{2}$, and the maximum density is $n-2$.  Given these bounds, we show that $\zeta_s$ and $\zeta_c$ are nearly as far apart as they can be.

Let $G$ be a graph on $n$ vertices with $m$ edges and sink of degree $d$.  Using a theorem of Merino Lopez \cite{MerinoLopez}, we can express the stationary density $\zeta_s(G)$ in terms of the Tutte polynomial $T(x,y)$ of $G$:
	\[ \zeta_s(G) = \frac1n \left( m-d + \frac{\frac{\partial T}{\partial y}(1,1)}{T(1,1)} \right) = \frac1n \left( m-d + \frac{u(G)}{\kappa(G)} \right). \]
Here~$u(G)$ is the number of spanning unicyclic subgraphs of~$G$, and $\kappa(G)$ is the number of spanning trees of~$G$.  In particular, for the complete graph $K_n$ it follows (see~\cite{wright}) that
	\[ \zeta_s(K_n) = \frac{n}{2} + \sqrt{\frac{\pi n}{8}} + o(n^{1/2}). \]
On the other hand, $\zeta_c(K_n) \geq n - 2 n^{1/2} \log n$: indeed, at this density, with high probability no sites in the initial configuration are unstable.
}

\section{Conclusions}

\old{
The conclusion of
\cite{absorbing5} that ``FES are shown to exhibit an absorbing state
transition with critical properties coinciding with those of the
corresponding sandpile model'' deserve to be re-evaluated.
One hope of the DMVZ paradigm was that critical features of the driven
dissipative model, such as the exponents governing the distribution of
avalanche sizes and decay of correlations, might be more easily
studied in the FES by examining the scaling behavior of these observables
as $\zeta \uparrow \zeta_c$.  However, the failure of the density conjecture,
and the continued aging of driven dissipative sandpiles beyond $\zeta_c$,
suggest that the two models may not share the same critical
features.
}

The conclusion of \cite{absorbing5} that ``FES are shown to exhibit an
absorbing state transition with critical properties coinciding with
those of the corresponding sandpile model'' should be
re-evaluated.

In response to this article, several researchers have suggested to us
that perhaps the density conjecture holds for stochastic sandpiles
even if not for deterministic ones.  This hypothesis deserves some
scrutiny.

For the driven dissipative sandpile, there is a
transition point at the threshold density of the FES, beyond which a
macroscopic amount of sand begins to dissipate.  The continued evolution
of the sandpile beyond $\zeta_c$ shows that driven sandpiles have (at least) a
one-parameter family of distinct critical states.  While the
stationary state has rightly been the object of intense study, we
suggest that these additional critical states deserve further
attention.
\\[-10pt]

\end{document}